# Giant Peltier Conductivity in an Uncompensated Semimetal $Ta_2PdSe_6$

**Akitoshi Nakano**[1*], **Ai Yamakage**[1*], **Urara Maruoka**[1], **Hiroki Taniguchi**[1], **Yukio Yasui**[2] **and Ichiro Terasaki**[1*]

[1] Department of Physics, Nagoya University, Nagoya 464-8602, Japan
[2] School of Science and Technology, Meiji Univ., 1-1-1 Higashi-Mita, Tama-ku, Kawasaki, Kanagawa 214-8571, Japan

**Abstract**

Thermoelectric properties of single crystal $Ta_2PdSe_6$ is investigated by means of transport measurements, and a density functional calculation. We found a giant Peltier conductivity of 100 $Acm^{-1}K^{-1}$ at 10 K and successfully explained it by means of conventional semiconductor theory. We concluded that an uncompensated semimetal, high mobility, and heavy effective mass are responsible for the giant Peltier conductivity. Our finding opens a new ground in the field of thermoelectrics to explore much better semimetals for a new possible application such as an electric current generator for a superconducting magnet.

**1. Introduction**

Itinerant electrons in solids in a thermal equilibrium are driven not only by an external electric field, but also by a temperature gradient $\nabla T$, and form a steady flow, i.e., an electrical current density $j$. While the former case is known as Ohm's law, the latter is rather recognized as the origin of the Seebeck effect that an electric field is generated by $\nabla T$ in an open circuit. In fact, Ohm used $j$ generated $\nabla T$ to find Ohm's law. The proportionality constant between $j$ and $\nabla T$ has been called the Peltier conductivity $P$ as $j = P(-\nabla T)$. Despite many studies for thermoelectrics, a theoretical upper limit of $P$ has yet to be explored, and accordingly an unexpectedly large $P$ has a great potential to revolutionize modern electronics.

Here, we report the observation of a giant $P$ of 100 Acm$^{-1}$K$^{-1}$ at 10 K in a single crystal of the layered semimetal Ta$_2$PdSe$_6$ [1]. This value is 200 times larger than the maximum $P$ of the commercially available thermoelectric material Bi$_2$Te$_3$ [2][3], and, to the best of our knowledge, it is the largest ever reported for a bulk material. This value may open a novel heat-to-electricity conversion such as a current generator, in which a 1 cc sample generates an electric current of 100 A with a temperature difference of 1 K. This is applicable to an isolated current source set in a cryogenic space. Using a two-carrier model for a perfectly uncompensated semimetal, we clarified why and how such a giant value is realized in the title compound.

## 2. Method

High-quality single crystals of Ta$_2$PdSe$_6$ were grown by means of I$_2$ vapor transport. Powders of tantalum (99.9%), palladium (99.9%), and selenium (99.9% or 99.999%) were loaded into an evacuated quartz tube with a I$_2$ concentration of ~3 mg/cm$^3$. Then, a temperature difference of 145 °C between 875 °C and 730 °C in a three-zone furnace was used for crystal growth for 4 days. Single phase polycrystalline sample was synthesized using the same starting powder. Once the raw powder was heated up to 550 °C, then the re-grinded powder was heated at 730 °C for 48 h in a tube furnace.

Ta$_2$PdSe$_6$ crystals were chemically characterized by scanning electron microscopy with energy dispersive X-ray spectroscopy (JEOL JSM-7500F). The ratio of Ta:Pd:Se was evaluated to be 1.91 : 0.93 : 6.14, which agrees with the stoichiometric ratio of Ta$_2$PdSe$_6$ (Fig. S1). We also conducted a synchrotron single crystal X-ray diffraction (XRD) measurement at BL02B1 in SPring-8. We employed a wavelength of 0.30963 Å to obtain a high-resolution data. We used a gas-blowing device for sample temperature control, and a Pilatus3 X 1M CdTe detector [4] for measuring two-dimensional (2D) diffraction patterns (Fig. S2). Diffraction intensity averaging was performed using *SORTAV*[5], and crystal structure refinement was performed by means of the *SHELXL* least squares program [6]. The refined ratio of Ta and Se against Pd was 1.997 and 5.998 respectively, indicating the atomic deficiency is less than 1%. A summary of the structural analysis is provided in Table 1 and S1.

Transport properties, including electrical resistivity, thermopower, and Hall resistivity, were measured using a PPMS (Quantum Design). The electrical resistivity was measured along the $b$-axis by a four-probe method using gold wires with 20 μm diameters and the silver paste. The thermopower along the $b$-axis was measured with a steady state and the two-probe technique. The sample bridged two separated copper heat baths, and the resistance heater (KYOWA KFLB-02-120-C1-11) created a temperature difference between the two heat baths, which was monitored through a copper-constantan differential thermocouple. The contribution of the voltage leads was carefully subtracted. The Hall resistivity with the four-probe technique was achieved by sweeping an out-of-plane magnetic field from - 4 to 4 T with a steady current along the $b$-axis. The typical setup for the transport measurements is shown in Fig. S3. The resistivity at each magnetic field were collected using ΔR mode of a nano-ohmmeter LR-700 (LINEAR RESEARCH INC). The Hall resistivity $\rho_{yx}$ was obtained by calculating $(\rho_{yx}(+H) - \rho_{yx}(-H))/2$. The specific heat measurements were performed by relaxation method with a commercial measurement system (Quantum Design PPMS Dynacool) by using polycrystalline $Ta_2PdSe_6$. The heat capacity was measured from 3 K to 300 K.

The band structure calculations were performed using the pseudopotential method based on the projector augmented wave (PAW) formalism [7] with plane-wave basis sets implemented in Quantum Espresso (version 6.6) [8]. The cut-off energies for plane waves and charge densities were set to 44 and 448 Ry in the full relativistic calculations. We conducted structure optimization using the structural parameter obtained from the single crystal XRD at 100 K as an initial input. We used a 14 × 14 × 4 uniform k-point mesh with the cold smearing method during self-consistent loops and 30 × 30 × 10 points with the "tetrahedra_opt" method for density-of-states and Fermi-surface calculation. Figure S4 shows the orbital-resolved Fermi surface.

**Result and Discussion**

The Peltier conductivity $P$ can be understood from the Seebeck effect, in which a voltage difference of $\Delta V$ is generated across a sample subjected to a temperature difference of $\Delta T$. The proportionality constant $S$ is called the Seebeck coefficient, and $\Delta V = S\Delta T$. In the presence of $\Delta T$, materials can behave like a battery with an open circuit voltage of $S\Delta T$ and an internal resistance of the sample $R$, as schematically shown in the inset of Fig. 1. In case the load resistance is much smaller than $R$, the maximum thermoelectric current is calculated to be $(S\Delta T)/R$, from which $P$ is evaluated as $S/\rho$ as a parameter intrinsic to materials, where $\rho$ is the resistivity. Figure 1 (a) shows $|P|$ of various materials [9] plotted as a function of conductivity $\sigma = 1/\rho$. We can find that $Ta_2PdSe_6$ locates at the top-level even at 300 K, and the highest at 10 K among the thermoelectric materials.

Figure 1 (b) shows a comparison of the temperature dependence of $|P|$ between $Ta_2PdSe_6$ and other representative low- and middle- temperature thermoelectric materials [10]-[14]. We find that $|P|$ of others except for $YbAgCu_4$ takes at most of the order of 1 $Acm^{-1}K^{-1}$, and gradually decreases as temperature decreases. Since optimized thermoelectric materials show a substantial residual resistivity accompanied by the $T$-linear $S$ at low temperatures, their $|P|$ is expected to be linear in $T$ at low temperatures. On contrary, $|P|$ of $Ta_2PdSe_6$ rapidly increases below 100 K to reach a giant value of 100 $Acm^{-1}K^{-1}$ at 10 K, indicating the giant $P$ results from an extremely low residual resistivity.

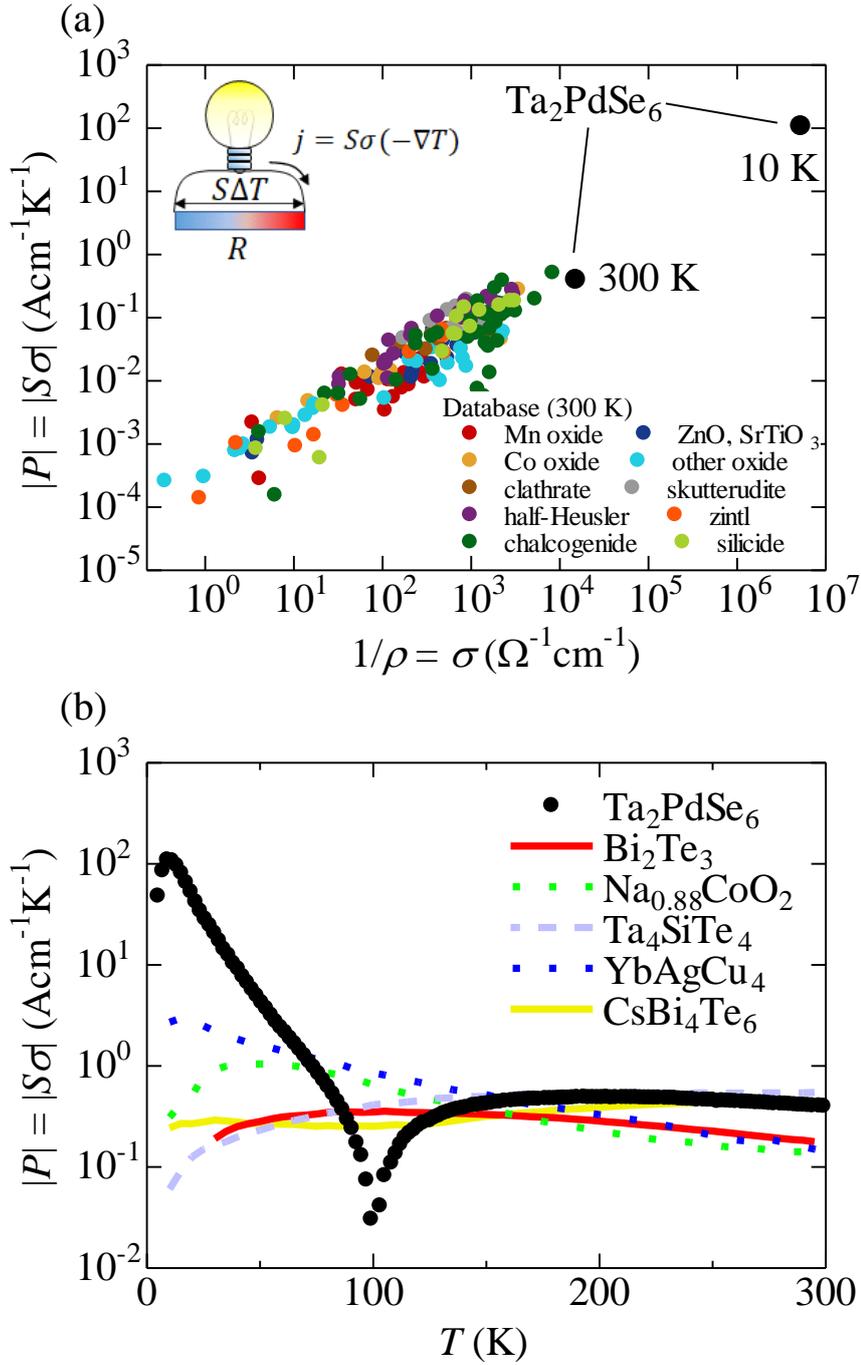

**Figure 1.** Comparison of the Peltier conductivity |P| between $Ta_2PdSe_6$ and other thermoelectric materials. (a) |P| for a wide variety of materials plotted against electrical conductivity ($1/\rho$). The data points except for $Ta_2PdSe_6$ are obtained from the Materials Research Laboratory database [9] of 300 K. (b) Temperature dependence of |P| for $Ta_2PdSe_6$, $Bi_2Te_3$ [10], $Na_{0.88}CoO_2$ [11], $Ta_4SiTe_4$ [12], $YbAgCu_4$ [13], and $CsBi_4Te_6$ [14].

We show $\rho$ and $S$ of single-crystal Ta$_2$PdSe$_6$ plotted as a function of temperature in Figures 2 (a) and 2 (b), respectively. The two quantities are measured along the longitudinal dimension (the $b$ axis shown in Figures 3 (a) and 3 (b)) of a needle-like single crystal, shown in the inset of Fig. 2 (a). The resistivity $\rho$ reaches a low value of 10$^{-7}$ $\Omega$cm at 2 K with a residual resistance ratio (rrr) of 694. This rrr value is much better than that of other chalcogenide semimetals [15][16]. $S$ takes a relatively large value of 40 µV K$^{-1}$ at 20 K. Consequently, the calculated $P$ result in the giant value of 100 Acm$^{-1}$K$^{-1}$ at 10 K as shown in Fig. 2 (c). Note that the power factor, which is a measure of electric power of the sample subjected to a temperature difference of 1 K, also becomes a huge value of 2.4 mWcm$^{-1}$K$^{-2}$ at 15 K, although we previously reported a relatively large value of 13 µWcm$^{-1}$K$^{-2}$ at 300 K [17]. We also pointed out this compound is semimetallic, for $S$ shows a sign change near 100 K. These trends are well reproduced between different single crystals (see Figs S5 (a) and S5 (b)).

We should point out that the contribution of the phonon drag effect to $S$ is negligible in Ta$_2$PdSe$_6$. As shown in Fig. S5, the single crystal prepared by using low-purity (99.9%) selenium powder shows worse conductivity than that prepared by high-purity (99.999%) selenium powder at the lowest temperature. This indicates that the carriers are scattered more frequently by the introduced impurities in the low-purity sample. Nevertheless, $S$ of the low-purity sample is almost the same as the high-purity one. If the phonon drag effect effectively contributed to $S$, $S$ would have to be affected by the impurity doping, since the mean free path of phonons is generally longer than that of electrons. This proves that the phonon drag effect is negligible and the diffusive part of electrons mainly contributes to $S$ of Ta$_2$PdSe$_6$.

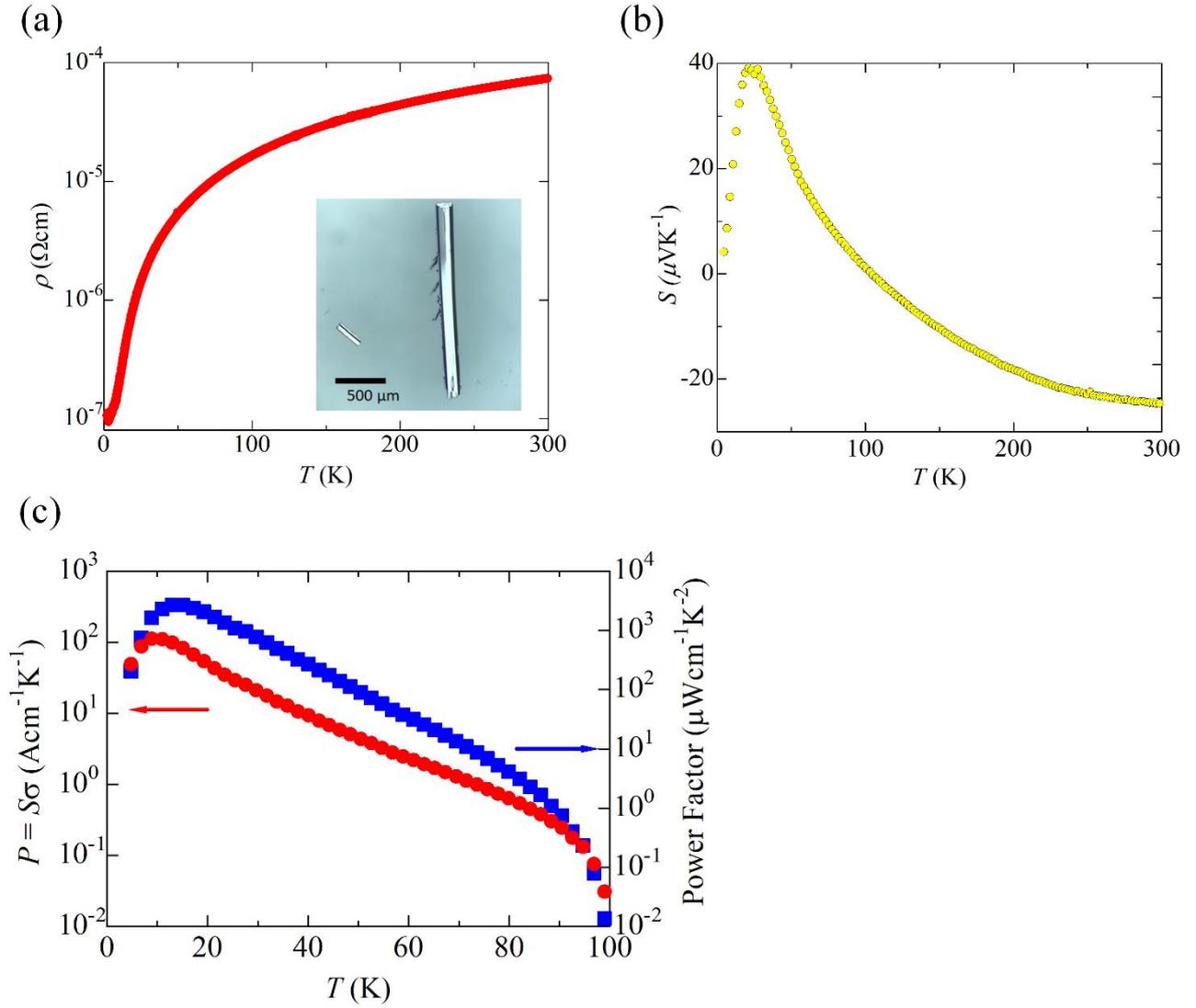

**Figure 2.** Electrical transport properties of $Ta_2PdSe_6$. (a) Temperature dependence of electrical resistivity ($\rho$), showing high conductivity at low temperature. The inset shows a photographic image of a single crystal. (b) Temperature dependence of thermopower ($S$). Note that the sign changes around 100 K. (c) Temperature dependence of the Peltier conductivity and power factor (= $S^2\sigma$).

Figures 3 (a) and 3 (b) shows the crystal structure of layered selenide $Ta_2PdSe_6$, the layers of which consist of face-shared $TaSe_6$ prisms and square-planar $PdSe_4$. The crystal structure is visualized by VESTA [18]. $Ta_2PdSe_6$ was first synthesized in 1985, and its electric resistivity was investigated [1]. $Ta_2PdSe_6$ has a structural similarity to the excitonic insulator candidate $Ta_2NiSe_5$ [19], a material in which we found unique structural [20][21] and transport properties [22][23]. Thus, we focused on $Ta_2PdSe_6$ as a related material and then identified the giant Peltier conductivity.

We now take a closer look at the electronic states of $Ta_2PdSe_6$. Before a band structure calculation, we performed a structural optimization to obtain a lowest-temperature structure by starting from a crystal structure determined by a synchrotron single-crystal X-ray diffraction measurement at 100 K (Table 1). The relaxed lattice and atomic coordination parameters are listed in Table 2. Figures 3 (c) and 3 (e) shows the Fermi surface visualized by FermiSurfer [24] and band dispersion around the Fermi energy ($E_F$) along the $MCLC_1$ path [25]. There is an electron Fermi surface near the I, X, and N points, while a hole Fermi surface near the $\Gamma$, Y, L, and Z points, indicating a semimetallic ground state. Figure 3 (f) shows the total and partial density of states, demonstrating that Ta $5d$ and Se $4p$ components mainly contribute to the low-energy electronic states. This is also shown by the orbital-resolved Fermi surface in Fig.S4. The calculated Fermi surface in Fig.3 (c) is shaped like a ragged, corrugated plane perpendicular to the $b$ axis, indicating a pseudo one-dimensional (1D) electronic structure.

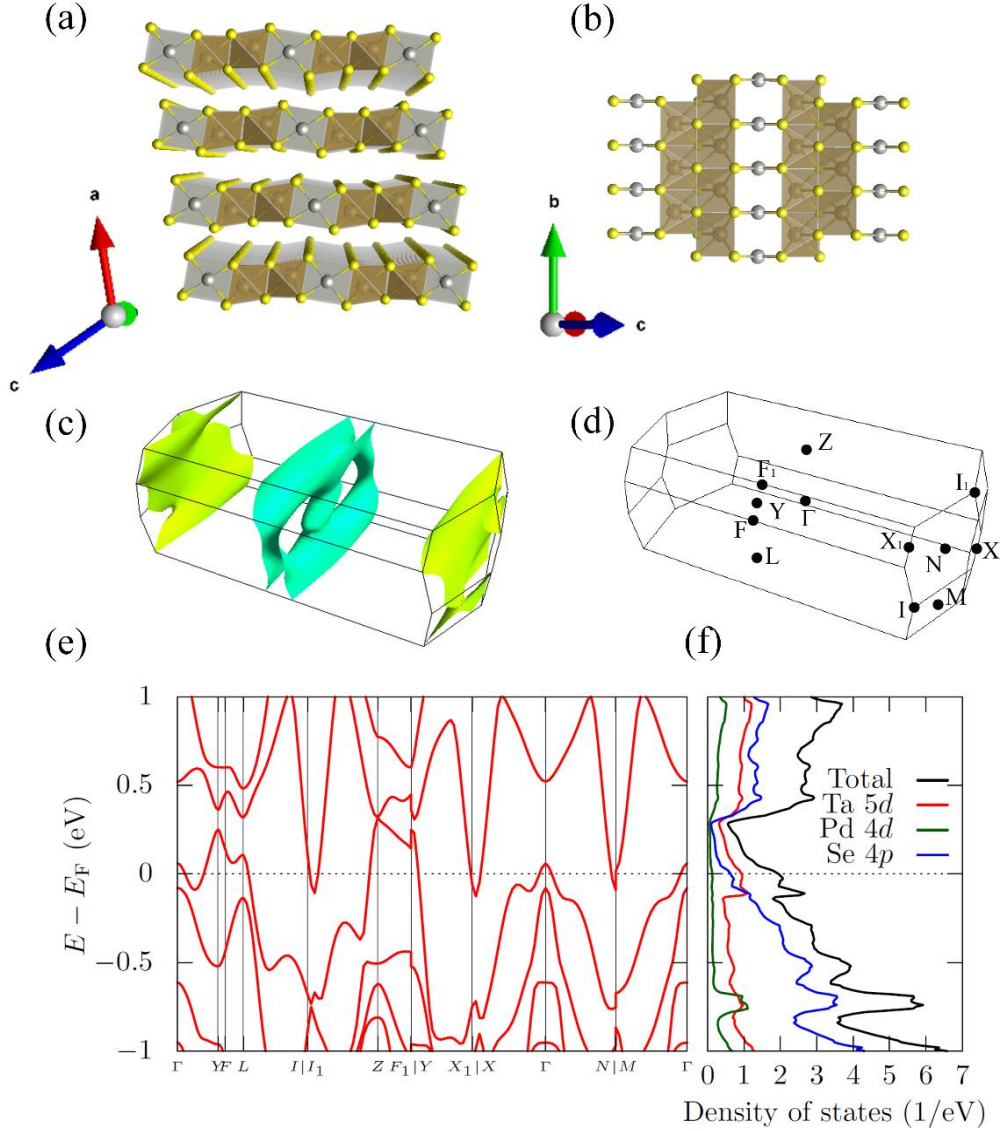

**Figure 3.** Crystal and electronic structure of Ta$_2$PdSe$_6$. (a) The layered crystal structure of Ta$_2$PdSe$_6$ viewed from the *b* axis. (b) The intra-layer structure viewed from the stacking direction (The visualization software is VESTA [18]). Each layer includes two types of one-dimensional chains, consisting of square-planar PdSe$_4$ and face-shared TaSe$_6$ prisms which stack and extend along the *b* axis. (c) Fermi surface of Ta$_2$PdSe$_6$ (The Fermi surface is visualized by FermiSurfer [24]). (d) High symmetry points of the Brillouin zone. (e) Band structure calculation along the MCLC$_1$ path, demonstrating a semimetallic ground state of Ta$_2$PdSe$_6$. (f) Density of state (DOS) around the Fermi energy ($E_F$). The partial DOS for Ta 5$d$, Pd 4$d$, and Se 4$p$ orbitals are also shown.

**Table 1.** Refined structural parameters for $Ta_2PdSe_6$ at 100 K determined by single crystal synchrotron X-ray diffraction. The space group is $C2/m$, and the obtained lattice parameters are $a = 12.4179(3)$ Å, $b = 3.3691(1)$ Å, $c = 10.3951(3)$ Å, and $\beta = 116.192(8)°$. The number of unique reflections for the refinement within the resolution limit d > 0.3 Å is 34192. The obtained $R$ factor and GOF are 3.36% and 1.047, respectively.

| Atom | Site | x | y | z | $B_{eq}(Å^2)$ |
|---|---|---|---|---|---|
| Ta | 4m | 0.67992(2) | 0 | 0.28974(2) | 0.1059(8) |
| Pd | 2a | 1/2 | 1/2 | 0 | 0.2258(16) |
| Se(1) | 4m | 0.71915(2) | 1/2 | 0.12513(2) | 0.1808(16) |
| Se(2) | 4m | 0.50566(2) | 1/2 | 0.23897(2) | 0.1713(16) |
| Se(3) | 4m | 0.84156(2) | 1/2 | 0.46069(2) | 0.1555(16) |

**Table 2.** Relaxed structural parameters for $Ta_2PdSe_6$. The space group is $C2/m$, and the lattice parameters are $a = 12.3296$ Å, $b = 3.34581$ Å, $c = 10.3233$ Å, and $\beta = 116.157°$.

| Atom | Site | x | y | z |
|---|---|---|---|---|
| Ta | 4m | 0.67830 | 0 | 0.28743 |
| Pd | 2a | 1/2 | 1/2 | 0 |
| Se(1) | 4m | 0.72160 | 1/2 | 0.12404 |
| Se(2) | 4m | 0.50348 | 1/2 | 0.24021 |
| Se(3) | 4m | 0.84163 | 1/2 | 0.46061 |

Now we explain how the giant $P$ realized in terms of a two-carrier model [26]. The partial conductivities of the electron ($\sigma_e$) and hole ($\sigma_h$) can be written as

$$\sigma_e = 2ne\mu_e, \quad (1)$$

$$\sigma_h = pe\mu_h, \quad (2)$$

where $n$ ($p$), $e$, and $\mu_{e(h)}$ are the concentration of electrons (holes) per valley, the element charge, and the carrier mobility of the electron (hole), respectively. Note that the factor of two in $\sigma_e$ represents the valley degeneracy. Then, the net Hall coefficient $R_H$ is described as

$$R_H = \frac{p\mu_h^2 - 2n\mu_e^2}{e(p\mu_h + 2n\mu_e)^2}. \quad (3)$$

Imposing a semimetallic condition of $2n = p$, we rewrite equation (3) as

$$R_H(f) = \frac{1}{pe}(2f - 1), \quad (4)$$

where $f = \mu_h/(\mu_h+\mu_e)$. For pseudo 1D parabolic bands, the partial thermopowers of electron ($S_e$) and hole ($S_h$) are given by (see supporting information)

$$S_e = -\alpha_0 m_e (p/2)^{-2} T, \quad (5)$$

$$S_h = \alpha_0 m_h p^{-2} T, \quad (6)$$

$$\alpha_0 = \frac{A^2 k_B^2}{4\pi^2 e h^2}, \quad (7)$$

where $k_B$, $h$, $m_{e(h)}$ and $A$, are the Boltzmann constant, the Planck constant, the effective mass of electrons (holes), and the cross-section of the Brillouin zone perpendicular to the $b$ axis, respectively. The net thermopower is given by

$$S = \frac{\sigma_e S_e + \sigma_h S_h}{\sigma_e + \sigma_h}. \quad (8)$$

Using equations (5)-(7), we get

$$S(f) = \alpha_0 p^{-2} T(f m_h - 4(1-f) m_e). \quad (9)$$

To find $p$, $m_h$, and $f$, we measured the Hall resistivity $\rho_{yx}$ at various temperatures, as shown in Fig. 4 (a). Little deviation from linear-field dependence implies that $R_H$ is well defined by $\rho_{yx}/\mu_0 H$ at 1 T. $R_H$ clearly shows a rapid decrease from $10^{-2}$ down to $10^{-3}$ cm$^3$C$^{-1}$ around 100 K, as shown in Fig. 4 (b). Assuming a heavily uncompensated condition of $f \sim 1$ at 2 K, we obtain $p \sim 7 \times 10^{20}$ cm$^{-3}$ from equation (4), which is roughly consistent with the calculated carrier concentration of $7.5 \times 10^{20}$ cm$^{-3}$. We also evaluate $\mu_h$ at 2 K to be $9 \times 10^4$ cm$^2$V$^{-1}$s$^{-1}$ from $R_H/\rho$. Furthermore, using equation (9), we get $m_h = 2.9 m_0$ ($m_0$ is the bare electron mass) for $S = 40$ $\mu$V K$^{-1}$ and $T = 20$ K. This value is roughly consistent with the effective mass of $3.5 m_0$ estimated from the electron specific heat coefficient $\gamma$ and $n$ (see Table 3 and Fig. S6). *Thus, we conclude that the lightly doped holes with high mobility and heavy mass are responsible for the giant Peltier conductivity in Ta$_2$PdSe$_6$.*

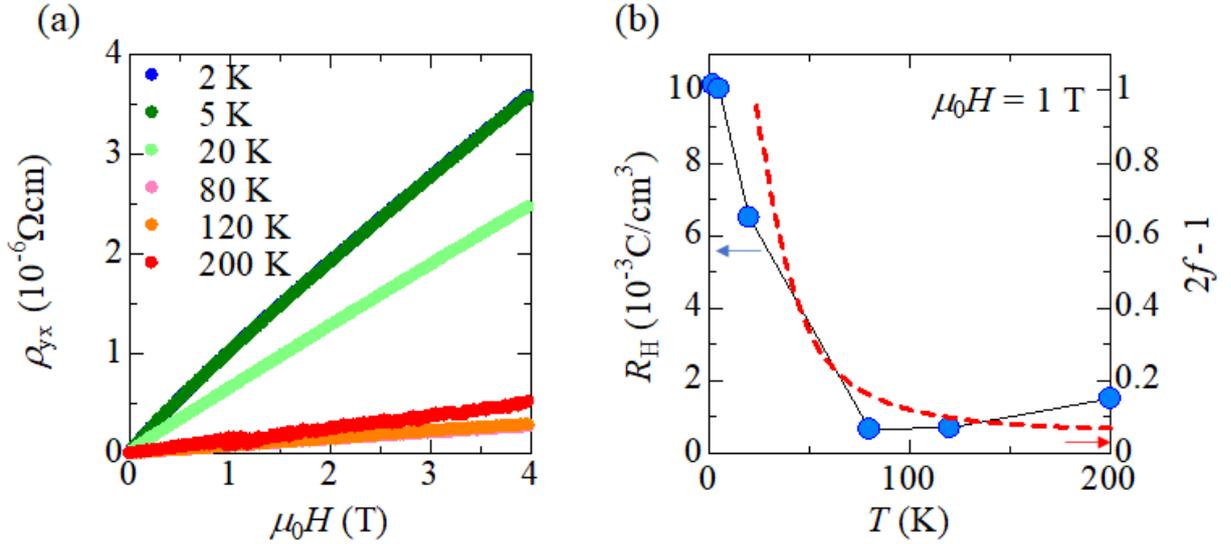

**Figure 4.** Magnetic transport properties of Ta$_2$PdSe$_6$. (a) Hall resistivity $\rho_{yx}$ at various temperatures. (b) Temperature dependence of Hall coefficient ($R_H$). The dotted line shows $2f-1$ calculated using the measured $S$ data (see text).

**Table 3.** Comparison of evaluated carrier concentration ($n$), electron specific coefficient ($\gamma$), and carrier effective mass ($m^*$) between the experiment and DFT calculation. $m_0$ is the bare electron mass

|  | Exp. | Calc. |
|---|---|---|
| $n$ (cm$^{-3}$) | 7×10$^{20}$ (hole) | 7.5×10$^{20}$ (hole, assuming $n = p$) |
| $\gamma$ (mJmol$^{-1}$K$^{-2}$) | 6.4 | 4.2 |
| $m^*$ | 2.9$m_0$ (Seebeck) | 0.4$m_0$ ($k \parallel b$) |
|  | 3.5$m_0$ (specific heat) | 2.2$m_0$ ($k \perp b$) |

In the field of thermoelectrics, the so-called B-factor is a measure of good thermoelectric materials [27]. It is proportional to $(m^*)^{3/2}\mu/\kappa_L$, where $m^*$, $\mu$, and $\kappa_L$ are the effective mass, the mobility, and the lattice thermal conductivity of a material, respectively. The numerator $(m^*)^{3/2}\mu$ characterizes the power factor and is 4000 times larger for Ta$_2$PdSe$_6$ than for Bi$_2$Te$_3$ ($m^* = 0.2m_0$ and $\mu = 1200$ cm$^2$V$^{-1}$s$^{-1}$). This is indicative of the large power factor in the present compound (2.4 mWcm$^{-1}$K$^{-2}$ at 15 K). The 1D electronic structure plays a vital role to achieve such high $(m^*)^{3/2}\mu$, since light holes along the 1D direction are responsible for the high mobility, whereas holes perpendicular to the 1D direction are responsible for the heavy mass [28]. We also note that Ta$_2$PdS$_6$, an isostructural compound of Ta$_2$PdSe$_6$, shows relatively high power factor ~ 30 µWcm$^{-1}$K$^{-2}$ at 300 K [17].

At higher temperatures, the electron conduction begins to contribute. Since we have determined $p$ and $m_h$ already, only $m_e$ is left as an unknown parameter in equation (9). Setting $m_e = 0.9m_0$ as an adjustable parameter, we obtain $f$ from the measured $S$ though equation (9). In Fig. 4 (b), we show the thus-obtained $2f$-1, which reasonably matches with $R_H(f)$. In short, all the measured transport parameters are quantitatively and consistently understood in terms of low carrier concentration, heavy mass, high hole mobility, and a crossover from $f = 1/2$ (compensated) to $f \sim 1$ (heavily uncompensated).

The giant Peltier conductivity and huge power factor can be used as a current source for a superconducting solenoid isolated in a cryogenic space. For 100 Acm$^{-1}$K$^{-1}$, a cubic sample of 1 cc would supply a thermoelectric current of 100 A to the zero-resistance solenoid for a temperature difference of 1 K. An absence of external current leads can make the system compact and concise and reduce cooling costs for such a solenoid. This can be a novel application of heat-to-electricity conversion. We notice that the induction voltage can be larger than $S\Delta T$, so that the field-sweeping rate must be kept low.

Our finding suggests that uncompensated semimetals can generate substantial electricity at low temperatures. This type of application will break new ground in the field of thermoelectrics. Semimetals are of high mobility in general and show good electrical conduction without impurity doping [29]. The uncompensated condition of $f \sim 1$ partly comes from electron-hole asymmetry as shown in Fig. 3 (c), and ternary or quaternary compounds may satisfy this condition. Such materials have never been researched, and therefore we believe that much better semimetals exist but are to date unknown.

**Acknowledgements**

Single-crystal XRD measurements were conducted with the approval of the Japan Synchrotron Radiation Research Institute (JASRI) (Proposal No. 2020A1431 and 2021A1159). This work was partly supported by Nanotechnology Platform Program (Molecule and Material Synthesis) of the Ministry of Education, Culture, Sports, Science and Technology (MEXT), Japan, Grant Number JPMXP09S20NU0029. This work is partially supported by Kakenhi Grant No. 19H05791, 20K20898, and 21K13878 of Japan.

# Supplementary Materials for

## Giant Peltier Conductivity in an Uncompensated Semimetal Ta$_2$PdSe$_6$

Akitoshi Nakano[1]*, Ai Yamakage[1]*, Urara, Maruoka[1], Hiroki Taniguchi[1], Yukio Yasui[2], Ichiro Terasaki[1]*

*Corresponding author. Email:
*nakano.akitoshi@nagoya-u.jp (A.N.)
*ai@st.phys.nagoya-u.ac.jp (A.Y.)
*terra@nagoya-u.jp (I.T.)

**This PDF file includes:**

Figs. S1 to S6

Table S1

Characterization of a single crystal Ta$_2$PdSe$_6$

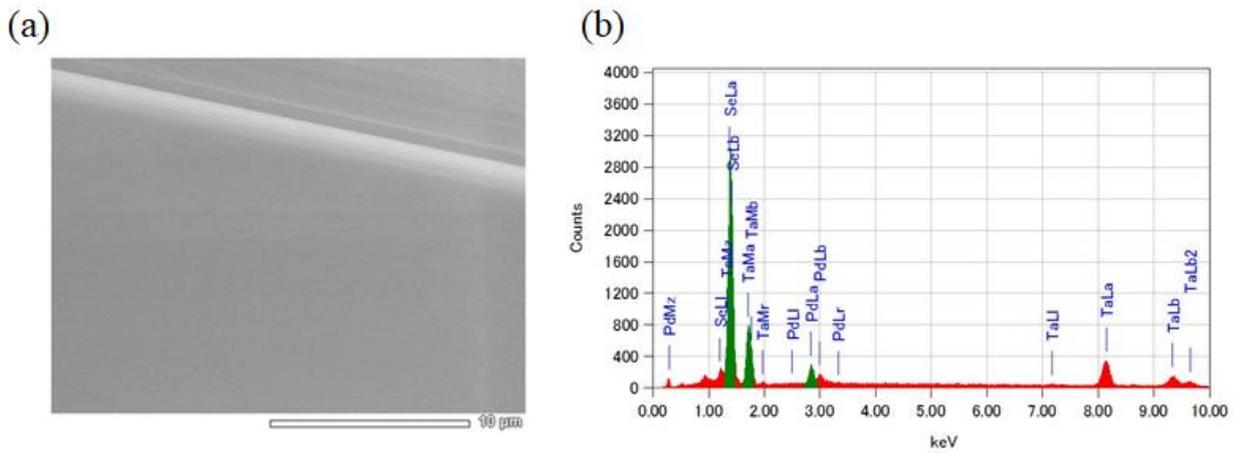

**Figure S1.** Characterization of a single crystal Ta$_2$PdSe$_6$. (a) A scanning electron microscope image, demonstrating a clean surface. (b) The result of energy dispersive X-ray spectroscopy. The ratio of Ta:Pd:Se is evaluated to be 1.912 : 0.93 : 6.14, which agrees with the stoichiometric ratio of Ta$_2$PdSe$_6$.

Synchrotron X-ray diffraction of a single crystal Ta$_2$PdSe$_6$

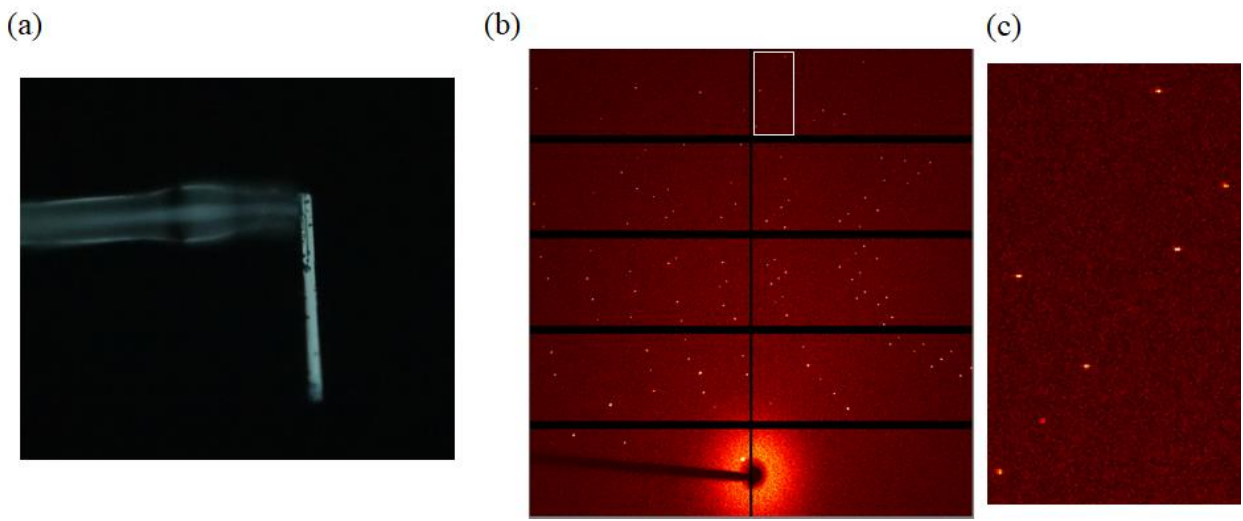

**Figure S2.** Synchrotron X-ray diffraction (XRD) of a single crystal Ta$_2$PdSe$_6$. (a) A single crystal used for measurements. The size of the crystal is 70 × 10 × 10 μm$^3$. (b) A XRD pattern of the single crystal Ta$_2$PdSe$_6$ measured at 100 K. The wave length is 0.30963 Å (c) The expanded XRD pattern at a high-resolution region (~0.3 Å$^{-1}$) surrounded by a white rectangular in b., demonstrating sharp reflections.

Structural analysis of Ta$_2$PdSe$_6$

**Table S1.**

| Chemical formula | Ta2 Pd Se6 |
|---|---|
| Temperature (K) | 100 |
| Wavelength (Å) | 0.30963 |
| Crystal dimension (μm$^3$) | 70 × 10 × 10 |
| Space group | *C2/m* |
| *a* (Å) | 12.4179(3) |
| *b* (Å) | 3.3691(1) |
| *c* (Å) | 10.3951(3) |
| *β* (°) | 116.192(8) |
| *V* (Å$^3$) | 390.25(3) |
| *Z* | 2 |
| *F*(000) | 1584 |
| (sin$\theta/\lambda$)$_{Max}$ (Å$^{-1}$) | 1.6665 |
| *N*$_{total,obs}$ | 34192 |
| *N*$_{unique,obs}$ | 7867 |
| Average redundancy | 4.3 |
| Completeness | 0.973 |
| wR$_2$ [# of reflections] | 0.0773 [6637] |
| R$_1$ [# of reflections] | 0.0336 [6637] |
| R$_1$ (I > 4σ)[# of reflections] | 0.0420 [5797] |
| GOF [# of reflections] | 1.047 |

Experimental setup for the transport measurements for Ta$_2$PdSe$_6$

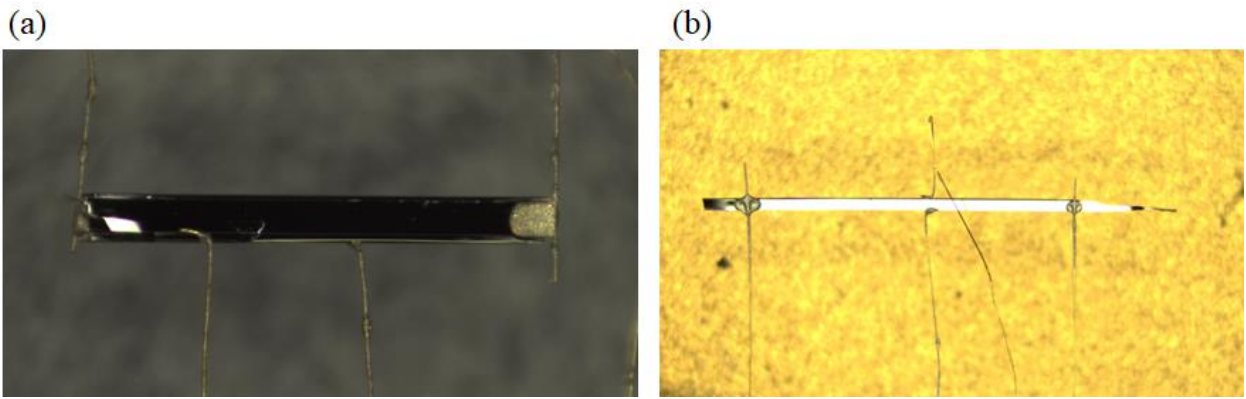

**Figure S3.** Experimental setup for the transport measurements for Ta$_2$PdSe$_6$. Four probe configurations for the resistivity (a) and Hall resistivity (b) measurements.  In the configuration of A, the four gold wires are attached to the side of the single crystal. The measured resistivity with this configuration corresponds to the data #7 in Fig. S3(a).

Orbital-resolved Fermi surface of a single crystal Ta$_2$PdSe$_6$

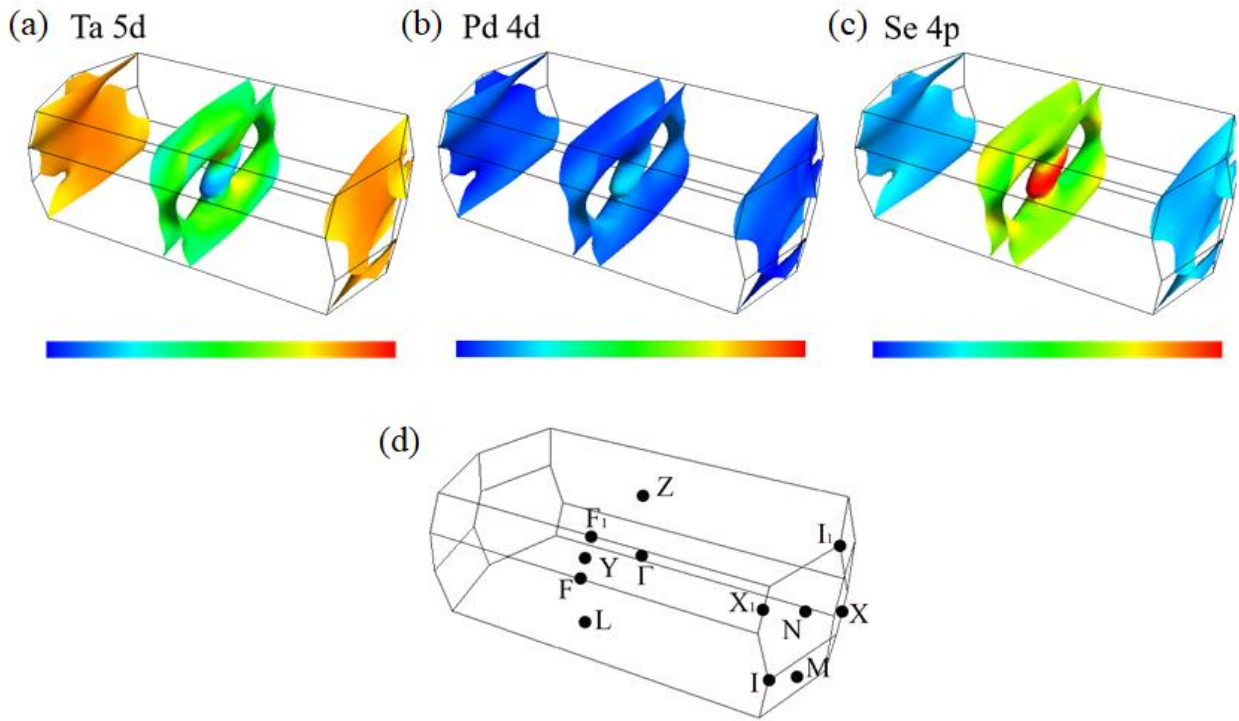

**Figure S4.** Orbital-resolved Fermi surface of a single crystal Ta$_2$PdSe$_6$. Calculated Fermi surfaces projected onto (a) Ta 5d, (b) Pd 4d, and (c) Se 4p orbitals. The Fermi surface is visualized by FermiSurfer. The electron and hole Fermi surfaces around the N and M points and those wrapping the Γ point consist of Ta 5d and Se 4p orbitals, respectively. On the hole Fermi surface near the Y, L, Z points, Ta 5d and Se 4p orbitals are strongly hybridized. (d) Time-reversal-invariant momenta (TRIM) on the Brillouin zone surface. The estimated carrier density $n$ and electronic specific heat coefficient $\gamma$ are $7.5 \times 10^{20}$ cm$^{-3}$ and 4.22 mJmol$^{-1}$K$^{-2}$, respectively.

Reproducibility of the transport measurements for $Ta_2PdSe_6$

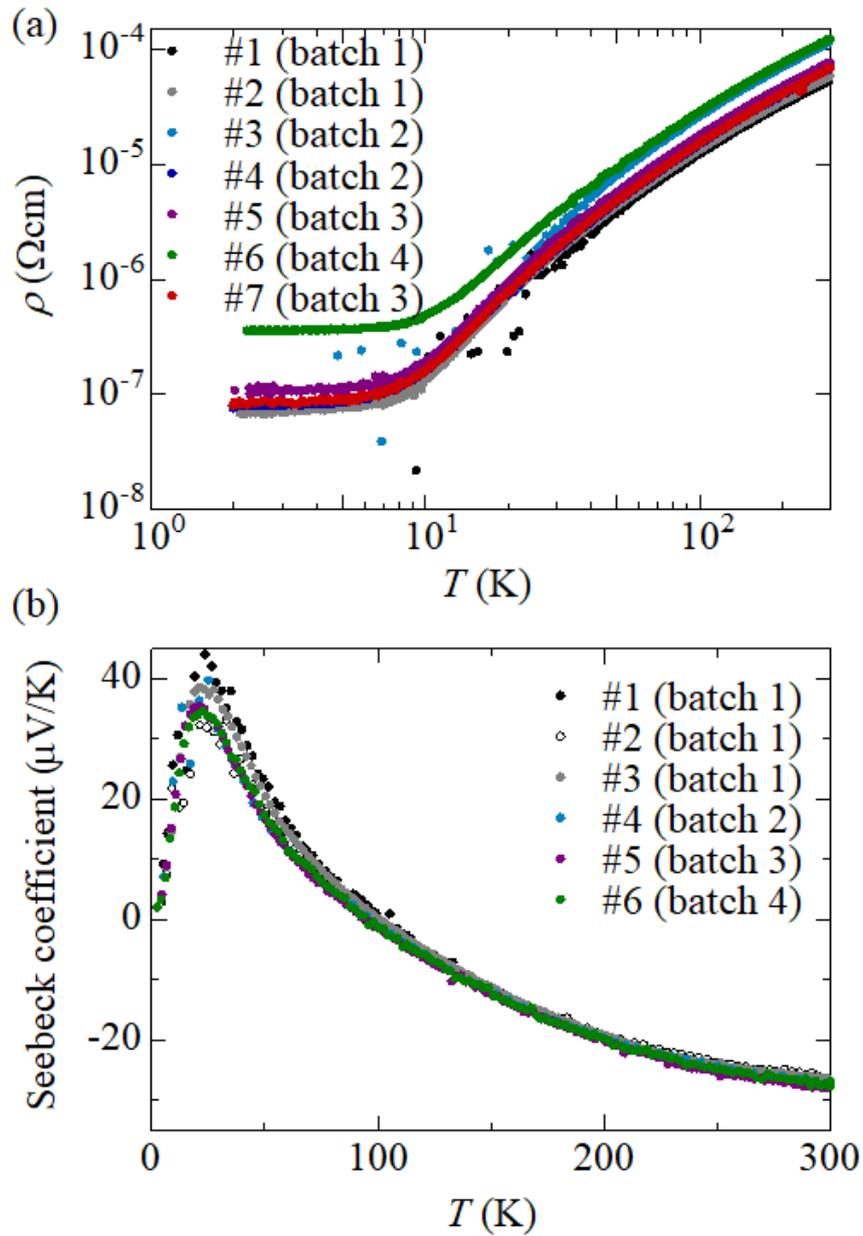

**Figure S5.** Reproducibility of the transport measurements for $Ta_2PdSe_6$. Temperature dependence of the (a) resistivity and (b) Seebeck coefficient for multiple single crystals of $Ta_2PdSe_6$ from different batches. Note that the purity of the Selenium powder of the batch 4 is 99.9 %, whereas that of batch 1~3 is 99.999%.

Derivation of the Seebeck coefficient for 1D parabolic bands

The Seebeck coefficient for metals is given by

$$S = \frac{\pi^2}{2} \frac{k_B}{e} \frac{k_B T}{E_f}, \tag{S1}$$

$$E_f = \frac{\hbar^2 k_f^2}{2m^*}. \tag{S2}$$

where $k_B, e, E_f, \hbar, k_f$ and $m^*$ are the Boltzmann constant, the elementary charge, the Fermi energy, the Dirac constant, the Fermi wave number, and the effective mass, respectively. Then the relationship between $k_f$ and the carrier density $n$ is expressed depending the dimensionality of the electronic state as

(3D) $$n = \frac{2}{(2\pi)^3} \frac{4\pi k_f^3}{3}, \tag{S3}$$

(2D) $$n = \frac{2}{(2\pi)^3} \pi k_f^2 \frac{2\pi}{c}, \tag{S4}$$

(1D) $$n = \frac{2}{(2\pi)^3} k_f \frac{2\pi}{b} \frac{2\pi}{c}, \tag{S5}$$

where, $b$ and $c$ are the lattice constant. Thus, using the equations (S1)~(S5), the Seebck coefficients depending on the dimensionality of the electronic state are given as

(3D) $$S = \frac{\pi^2 k_B^2}{e\hbar^2} (3\pi^2 n)^{-\frac{2}{3}} m^* T, \tag{S6}$$

(2D) $$S = \frac{\pi k_B^2}{2e\hbar^2 c} \frac{m^*}{n} T, \tag{S7}$$

(1D) $$S = \frac{k_B^2}{e\hbar^2 b^2 c^2} \frac{m^*}{n^2} T. \tag{S8}$$

In the main text, we use the cross-section of the Brillouin zone instead of $\frac{2\pi}{b} \times \frac{2\pi}{c}$ for the equation S8.

Specific heat measurement for Ta$_2$PdSe$_6$

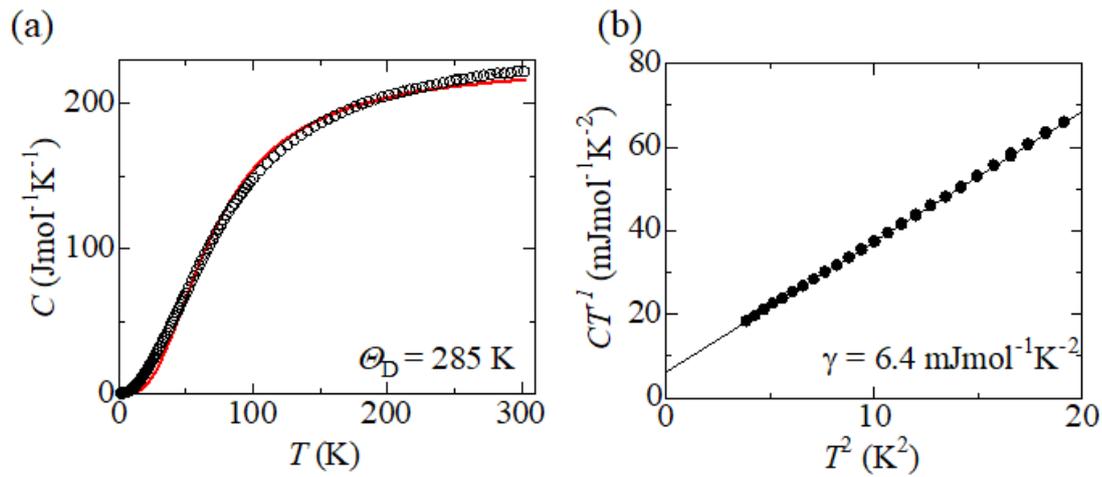

**Figure S6.** Specific heat measurement for Ta$_2$PdSe$_6$ (a)Temperature dependence of the specific heat (*C*). The red line shows the fitting to the experimental data by using the Debye model. The estimated Debye temperature ($\Theta_D$) is 285 K. (b) $CT^{-1}$ vs $T^2$ plot. The black line shows the fitting to the experimental data by assuming $C = \gamma T + AT^3$, where $\gamma$ and $A$ are electronic and lattice specific heat coefficient, respectively. The effective mass from $\gamma$ is estimated to be 3.5$m_0$ using the free electron model.